\RequirePackage{ifpdf}
\documentclass[12pt,letterpaper]{article}
\usepackage{jheppub}
\usepackage{epsfig}
\usepackage[latin1]{inputenc}
\usepackage{bbm,amsfonts}
\usepackage{graphicx}
\usepackage{amssymb,amsmath,amsfonts,mathtools}
\usepackage{fancybox}
\usepackage{enumerate}
\usepackage{dsfont}
\usepackage{verbatim}
\usepackage{wrapfig}
\usepackage{slashed}

\author[a]{Marco S. Bianchi}
\author[b]{and Andrea Mauri}
 
\affiliation[a]{Center for Research in String Theory - School of Physics and Astronomy Queen Mary University of London, Mile End Road, London E1 4NS, UK}
\affiliation[b]{Dipartimento di Fisica, Universit\`a degli Studi di Milano-Bicocca and INFN, Sezione di Milano-Bicocca, Piazza della Scienza 3, I-20126 Milano, Italy}

\emailAdd{m.s.bianchi@qmul.ac.uk}   
\emailAdd{andrea.mauri@mi.infn.it} 
\preprint{QMUL-PH-17-20}
  
\abstract{We consider the Bremsstrahlung function associated to a 1/6-BPS Wilson loop in ABJM theory, with a cusp in the couplings to scalar fields. We non-trivially extend its recent four-loop computation at weak coupling to include non-planar corrections. We have recently proposed a conjecture relating this object to supersymmetric circular Wilson loops with multiple windings, which can be computed via localization. We find agreement between this proposal and the perturbative computation of the Bremsstrahlung function, including color sub-leading corrections. This supports the conjecture and hints at its validity beyond the planar approximation.
}

\title{ABJM $\theta$-Bremsstrahlung at four loops and beyond: non-planar corrections}

\keywords{ABJM theory, BPS Wilson loops, Cusp, Bremsstrahlung function}

\newcommand{\be}{\begin{equation}}
\newcommand{\ee}{\end{equation}}
\newcommand{\beq}{\begin{equation}}
\newcommand{\eeq}{\end{equation}}
\newcommand{\bea}{\begin{eqnarray}}
\newcommand{\eea}{\end{eqnarray}}
\newcommand{\ena}{\end{eqnarray}}

\def\Tr{\textrm{Tr}}

\numberwithin{equation}{section}

\def\clock{{\count0=\time
           \divide\count0 60
           \ifnum\count0<10 0\fi\the\count0
           \multiply\count0 -60 \advance\count0 \time
           :\ifnum\count0<10 0\fi \the\count0
         }}
\newcommand{\timestamp}{{\small\vbox{\hbox{\tt\jobname.tex}
\hbox{\the\day/\the\month/\the\year, \clock}}}}
\begin{document}

\maketitle
\allowdisplaybreaks

\section{Introduction}

ABJM theory in three dimensions \cite{Aharony:2008ug,Aharony:2008gk} can be localized \cite{Pestun:2007rz}, which allows for computing some supersymmetric observables exactly, or at least boiling their computation down to a matrix model average \cite{Kapustin:2009kz,Marino:2009jd,Drukker:2010nc}. This in particular includes the expectation value of certain supersymmetric circular Wilson loops.
Remarkably, also the calculation of specific non-BPS quantities can be related to such objects and therefore mapped to localization results. This reasoning applies (with a certain degree of speculation) to the small angle limits of generalized cusps constructed with supersymmetric Wilson lines, i.~e.~the Bremsstrahlung functions, and another example is the limit considered in \cite{Bianchi:2016rub}.

In this paper we focus on the cusp constructed with two 1/6-BPS Wilson lines  meeting at a geometric angle $\varphi$ whose connection is endowed with a coupling to the scalars of the theory \cite{Berenstein:2008dc,Drukker:2008zx,Chen:2008bp,Rey:2008bh}.
It is possible to introduce a second angle $\theta$, describing a kick in the coupling to the scalars of the theory, occurring at the same Wilson loop time as the geometric cusp \cite{Griguolo:2012iq}.
Both angles produce a divergence and the resulting cusp anomalous dimension depends on two angles.
The small angle limits of this object are controlled by Bremsstrahlung functions and in this case we distinguish that associated to the geometric angle $B_{1/6}^{\varphi}$ and that associated to the internal space one $B_{1/6}^{\theta}$, which are a priori two different objects.

In a recent publication \cite{Bianchi:2017afp} we computed the weak coupling four-loop corrections to the $\theta$-Bremsstrahlung function (see \cite{Grozin:2017css} for a recent four-loop computation in four-dimensional QCD). Based on this result, we put forward a conjecture that relates $B_{1/6}^{\theta}$ to $B_{1/6}^{\varphi}$ by a simple factor of 2. The geometric angle Bremsstrahlung function $B_{1/6}^{\varphi}$, in turn, is conjectured \cite{Lewkowycz:2013laa} to be computable in terms of supersymmetric 1/6-BPS Wilson loops with multiple windings, which can be evaluated \cite{Marino:2009jd,Drukker:2010nc,Klemm:2012ii} using localization \cite{Pestun:2007rz} in the ABJM model \cite{Kapustin:2009kz}.
Both the original analysis of \cite{Lewkowycz:2013laa} and the computation and consequent conjecture of \cite{Bianchi:2017afp} were confined to the planar approximation. In this paper we non-trivially extend that computation to color sub-leading corrections and widely comment on the result we obtain.

The paper is structured as follows.
In section \ref{sec:cusp} we begin by reviewing the basics of the cusp constructed with two 1/6-BPS rays and the conjecture for the $\theta$-Bremsstrahlung function. Next, in section \ref{sec:computation}, we outline its calculation at four loops in weak coupling perturbation theory. We perform the computation for generic gauge group ranks $N_1$ and $N_2$. In particular, no planar approximation is enforced. The steps of the computation are qualitatively similar to the analysis in \cite{Bianchi:2017afp} and we mostly stress the new features of the color sub-leading corrections. More technical details can be found in appendix \ref{app:masters}, where specifically all the required master integrals are defined and evaluated.
Comparison with a localization based prediction for the Bremsstrahlung function is attained by first determining an explicit expression for the 1/6-BPS circular Wilson loop, multiply wound around the great circle. We perform this starting from the matrix model \cite{Kapustin:2009kz,Marino:2009jd,Drukker:2010nc} obtained from localization of the ABJM model. In particular, in section \ref{sec:matrixmodel} we expand the matrix model average for the Wilson loop at weak coupling up to the required order for comparison with the perturbative computation of the Bremsstrahlung function, that is four loops and for generic gauge group ranks, including the complete genus expansion (which at four loops entails both sub-leading $\sim 1/N^2$ and sub-sub-leading $\sim 1/N^4$ corrections).
Finally, we compare the predicted result for the $\theta$-Bremsstrahlung function stemming from the conjecture and our perturbative computation and observe perfect agreement.
We interpret this as a strong evidence for the correctness of the proposal of \cite{Bianchi:2017afp} and for its extension beyond the planar limit.

\section{The ABJM $\theta$-Bremsstrahlung}\label{sec:cusp}

We consider the $U(N_1)_k\times U(N_2)_{-k}$  ABJ(M) model. This is a level $k$ Chern-Simons theory, described in terms of the two gauge fields $A$ and $\hat{A}$ which couple to a matter sector given by bi-fundamental complex scalars $C_I, \bar C^J$ and fermions $\psi^I, \bar \psi_J$ (with $I,J = 1,\dots 4$). We work with the Euclidean version of the ABJM action and refer the reader to \cite{Bianchi:2017svd} for the its full expression and corresponding Feynman rules. 

The 1/6-BPS Wilson loop \cite{Drukker:2008zx,Chen:2008bp,Rey:2008bh} 
\begin{equation}\label{eq:WL}
W_{1/6}[\Gamma] = \frac{1}{N_1}\, \Tr \left[ \,\textrm{P}\exp{ -i \int_\Gamma d\tau~ \left( A_{\mu} \dot x^{\mu}-\frac{2 \pi i}{k} |\dot x|\, M_{J}^{\ \ I} C_{I}\bar C^{J} \right)(\tau) } \right] 
\end{equation}
can be defined on a cusped contour $\Gamma$ 
\begin{equation}
\Gamma: \ \ \ \ x^{0}=0 \ \ \ \ \ x^{1}=s\cos\frac{\varphi}{2} \ \ \ \ \  x^{2}=|s|\sin\frac{\varphi}{2}\ \ \ \ \  -\infty\le s\le \infty
\end{equation}
given in terms of the angle $\varphi$. The geometric cusp can be generalized by the introduction of an additional internal angle $\theta$,  by taking different scalars coupling matrices $M$ on the two edges of the cusp 
\begin{equation} \label{eq:Mmatrices}
M_{1J}^{\ \ I}=\mbox{\small $\left(
\begin{array}{cccc}
 -\cos \frac{\theta }{2}& -\sin \frac{\theta }{2} & 0 & 0 \\
 -\sin \frac{\theta }{2}& \cos\frac{\theta }{2} & 0 & 0 \\
 0 & 0 & -1 & 0 \\
 0 & 0 & 0 & 1
\end{array}
\right)$}
\ \ \ \ \mathrm{and}\ \ \ \  M_{2J}^{\ \ I}=\mbox{\small $\left(
\begin{array}{cccc}
 -\cos \frac{\theta }{2} & \sin \frac{\theta }{2} & 0 & 0 \\
 \sin \frac{\theta }{2} & \cos\frac{\theta }{2} & 0 & 0 \\
 0 & 0 & -1 & 0 \\
 0 & 0 & 0 & 1
\end{array}
\right)$}
\end{equation}
Unlike the standard straight line case, for this configuration no BPS condition holds leading to an unprotected operator which develops an anomalous dimension  $\Gamma_{1/6}(\varphi,\theta)$ that depends on the two angles and the couplings of the model. 
The small angles limits of $\Gamma_{1/6}(\varphi,\theta)$ are controlled by corresponding Bremsstrahlung functions
$$
\Gamma_{1/6}(\varphi,\theta) \sim \theta^2 B^{\theta}_{1/6}-\varphi^2 B^{\varphi}_{1/6}
$$
which are in principle two independent functions of the couplings. Quite surprisingly  the weak coupling expansions of these observables suggest that they might be related in a simple fashion. Indeed,  the geometric angle Bremsstrahlung $B^{\varphi}_{1/6}$ has been directly computed up to two loops  \cite{Griguolo:2012iq,Bianchi:2014laa} and a conjecture for its all-loop expression has been given in \cite{Lewkowycz:2013laa, Bianchi:2017afp}. This also agrees with the strong coupling result up to the sub-leading order \cite{Correa:2014aga,Aguilera-Damia:2014bqa}.  

The internal angle  Bremsstrahlung function $B_{1/6}^{\theta}(k,N)$, has been computed at weak coupling at two- \cite{Bianchi:2014laa} and four-loop orders in the planar limit \cite{Bianchi:2017afp}. It was found that  the simple relation 
\begin{align} \label{bremconj}
2\, B_{1/6}^{\theta}(k,N) & \underset{\underset{N_{1,2}\gg 1}{\text{conj \cite{Bianchi:2017afp}}}}{=} B_{1/6}^{\varphi}(k,N) \underset{\underset{N_{1,2}\gg 1}{\text{conj \cite{Lewkowycz:2013laa}}}}{=} \frac{1}{4\pi^2}\, \partial_n\, |W_n(k,N_1,N_2)|\, \bigg|_{n=1} 
\end{align}
holds up to the four-loop order and in the planar approximation. The simplicity of this relation, together with some additional group theoretical  arguments \cite{Bianchi:2017afp}, suggests that it could hold true at all loops.  

In the absence of a general argument justifying the relation (\ref{bremconj}), further checks are needed to avert the possibility that it could be an accident of the first few perturbative orders.  In this paper we provide a  striking test of (\ref{bremconj}) by computing the full non-planar $B_{1/6}^{\theta}(k,N)$ at four loops and comparing it with the conjectural form of  $B_{1/6}^{\varphi}(k,N)$. 

\section{The perturbative computation}\label{sec:computation}

\begin{figure}[h]
\centering
\includegraphics[scale=0.325]{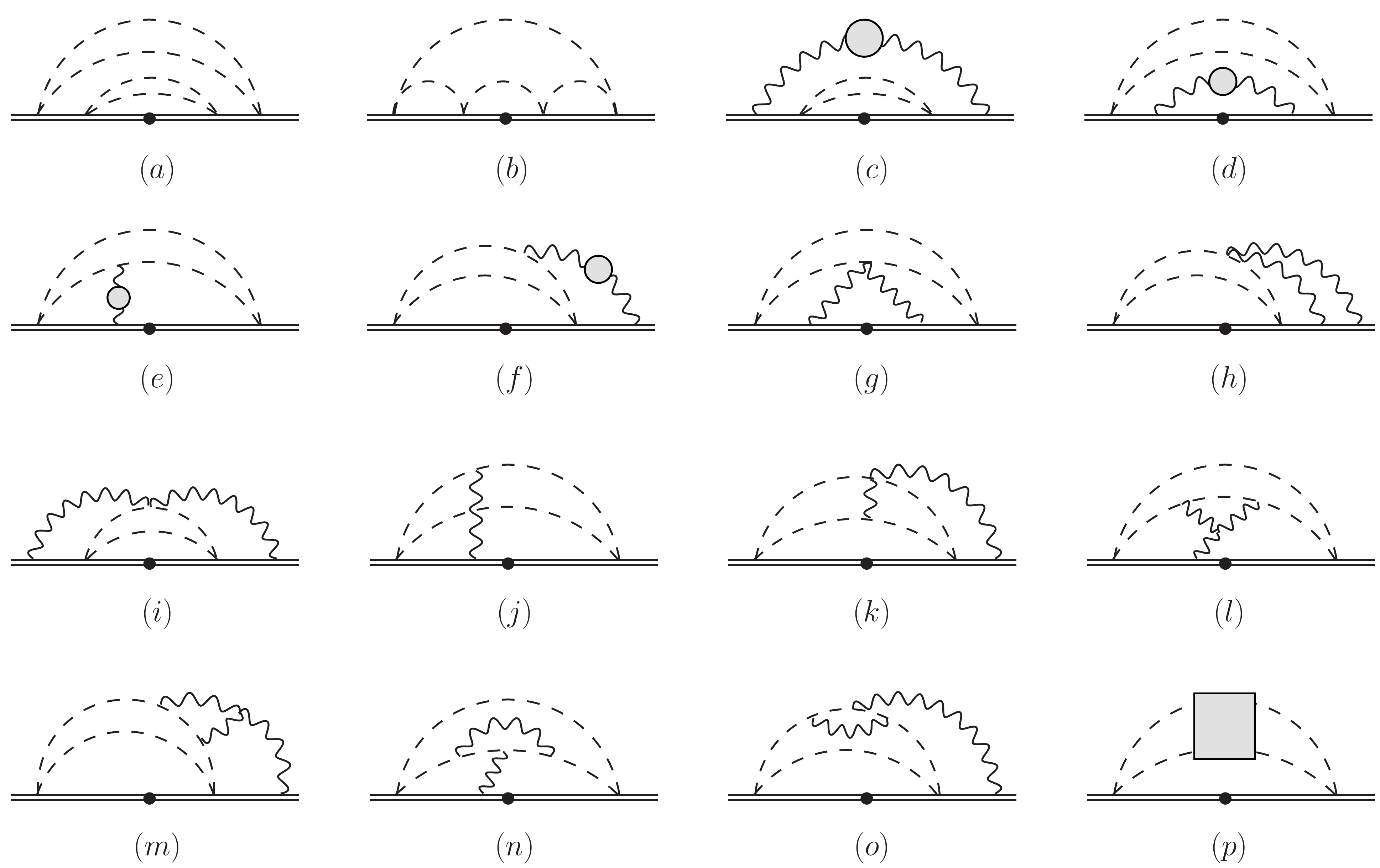}
\caption{List of planar diagrams contributing to the four-loop $\theta$-Bremsstrahlung function. Gray bullets stand for 1-loop corrections to the gauge propagator. The gray box collects the two-loop corrections to the bi-scalar two-point function.} \label{fig:4loops}
\end{figure}

\begin{figure}[h]
\centering
\includegraphics[scale=0.35]{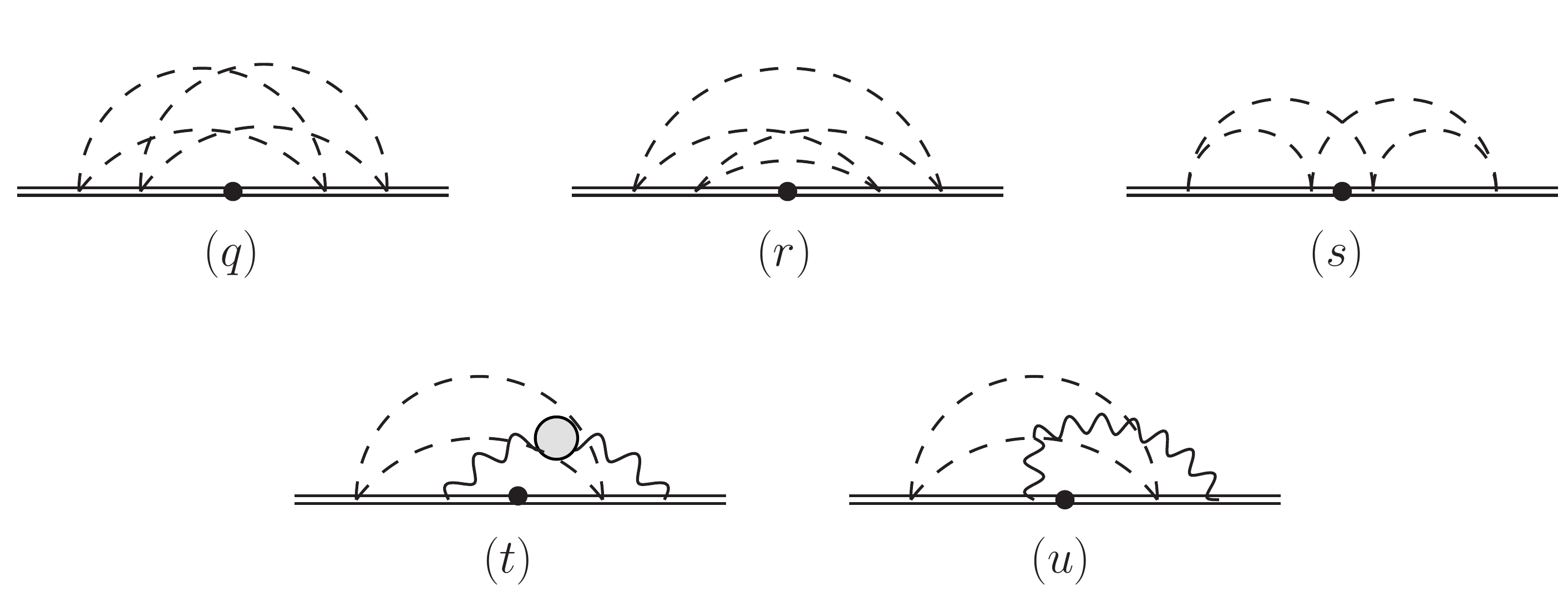}
\caption{List of non-planar diagrams contributing to the four-loop $\theta$-Bremsstrahlung function.} \label{fig:4loopsNP}
\end{figure}

The computation of $B_{1/6}^{\theta}(k,N)$ at finite $N$ goes along the lines of its planar counterpart and we refer the reader to \cite{Bianchi:2017afp} for further details.  As a result of the analysis of  \cite{Bianchi:2017afp}, the calculation can be performed taking advantage of a number of relevant simplifications. Indeed, we only need to consider 1PI four-loop diagrams at $\varphi=0$ with relevant factors of the angle $\theta$ for the $\theta$-Bremsstrahlung function. In particular this requires the presence of at least two insertions of the scalar bilinear operators in the connection \eqref{eq:WL}.
These diagrams are depicted in Figure \ref{fig:4loops} and  \ref{fig:4loopsNP}.
We use standard notation to represent the fields, with double lines for the Wilson contours,  solid, curly and dashed lines for fermion, vector and scalar fields respectively.
Diagrams (a)-(o) are topologically planar and are the same as in \cite{Bianchi:2017afp}. Some of them have contractions which generate color sub-leading terms, which we have computed generalizing the analysis in  \cite{Bianchi:2017afp}. These can arise by exactly the same diagrams as in the planar case, but where different possible contractions of the fields are possible, or by diagrams which are topologically the same but where different gauge fields contribute. For instance, in the diagrams with a one-loop gluon self-energy, also the mixed $\langle A \hat A\rangle$ has to be considered.

\begin{figure}[h]
\centering
\includegraphics[scale=0.35]{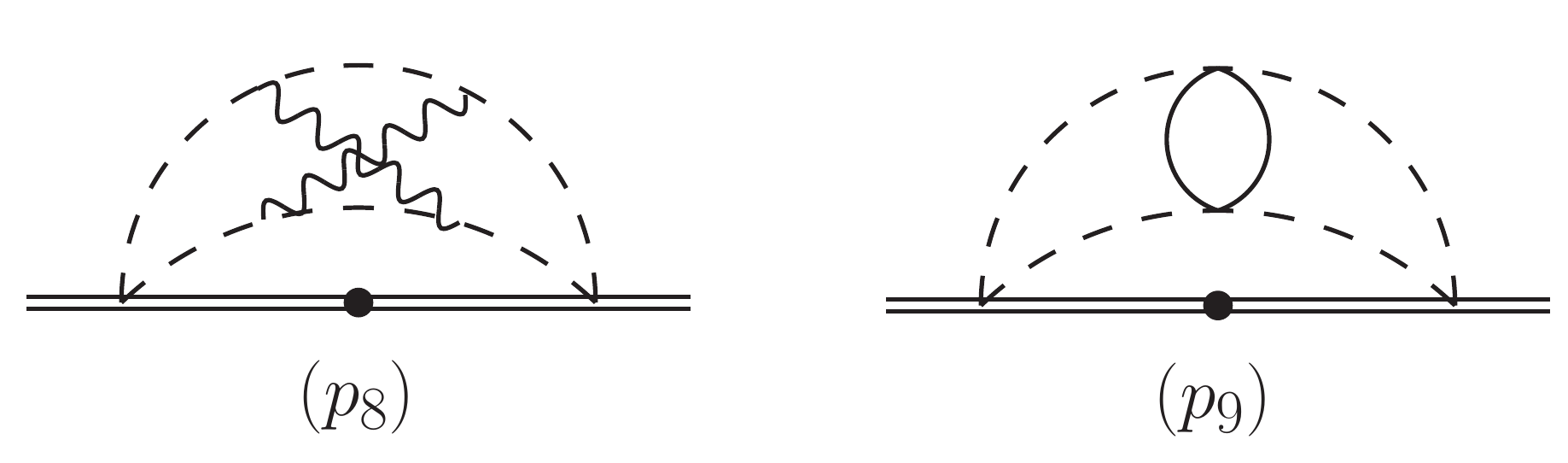}
\caption{non-planar corrections to the scalar bilinear two-point function.} \label{fig:pNP}
\end{figure}

Diagram (p) collects the corrections to the scalar bilinear two-point function. These include planar topologies as in  \cite{Bianchi:2017afp} and two new non-planar diagrams, which we depict in Figure \ref{fig:pNP}. Incidentally, these two diagrams are both proportional to $(N_1-N_2)$ and hence do not contribute for equal gauge group ranks. The detailed  results for diagrams (p) are reported in appendix \ref{app:vertex}. 

Diagrams (q)-(u) are genuinely non-planar and need to be evaluated from scratch. 
We recall that in some graphs the cusp point can be placed in different inequivalent positions along the Wilson line. We sum over all these configurations, provided they generate $\theta$-dependent factors.
A bunch of other possible diagrams are found to vanish and we have not displayed them here. For instance these include the non-planar version of diagrams (l) and (m) where the gluons land on different scalar propagators. Moreover all graphs with two separate gluon propagators connecting a scalar and the Wilson line (there are 8 topologically inequivalent such graphs) are found to vanish. This can be argued solely on (anti)symmetry grounds.

The computational steps are the same as those followed in the planar case  \cite{Bianchi:2017afp}.  We first write the diagrams in momentum space and convert contour integrations to heavy quark effective theory  (HQET) propagators \cite{Grozin:2015kna,Bianchi:2017svd}. Then we perform reduction to master integrals using  LiteRed \cite{Lee:2012cn,Lee:2013mka} and FIRE \cite{Smirnov:2008iw,Smirnov:2013dia,Smirnov:2014hma}, which  implement the application of integration by parts identities (IBP) \cite{Tkachov:1981wb,Chetyrkin:1981qh}. 

Graphs (q) and (t) have a central HQET propagator which is a linear combination of the other two. Thanks to the linearity of HQET propagators they can be decomposed by partial fractioning into planar integral topologies, which we performed by an automated routine.
Diagrams (r), (s) and (u) involve instead new non-planar master integrals.

After reduction each diagram is expressed as a linear combination on a basis of 21 planar master integrals which have been fully dealt with in \cite{Bianchi:2017afp} and three additional non-planar integrals. The full list of master integrals is sketched in Figure \ref{fig:master} and they are explicitly defined and evaluated in appendix \ref{app:masters}.
\begin{figure}
\begin{center}
\includegraphics[scale=0.4]{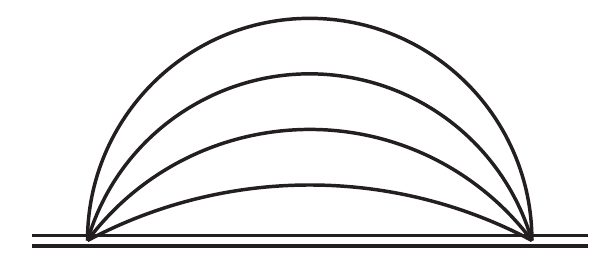} \hspace{2mm}
\includegraphics[scale=0.4]{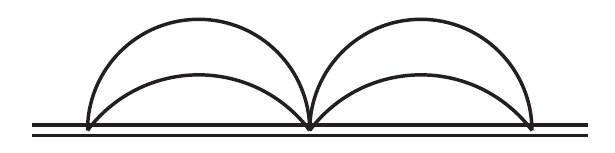} \hspace{2mm}
\includegraphics[scale=0.4]{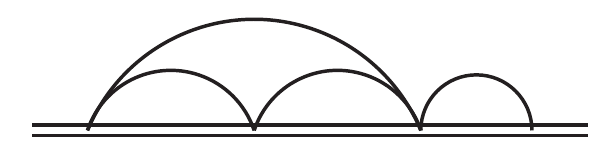} \hspace{2mm}
\includegraphics[scale=0.4]{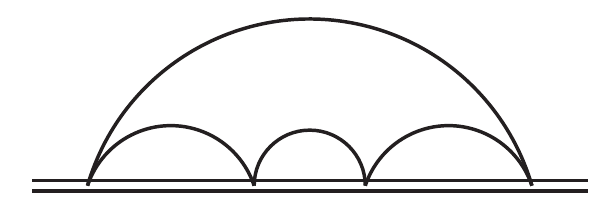} \\[2mm]
\includegraphics[scale=0.4]{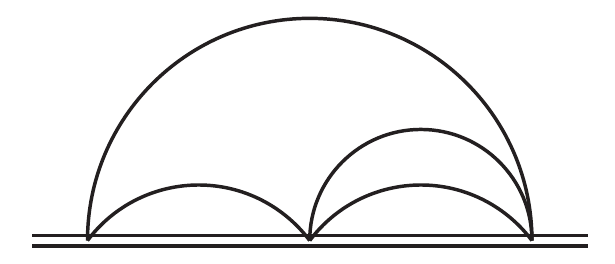} \hspace{2mm}
\includegraphics[scale=0.4]{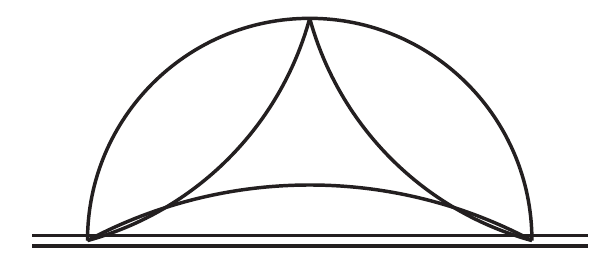} \hspace{2mm}
\includegraphics[scale=0.4]{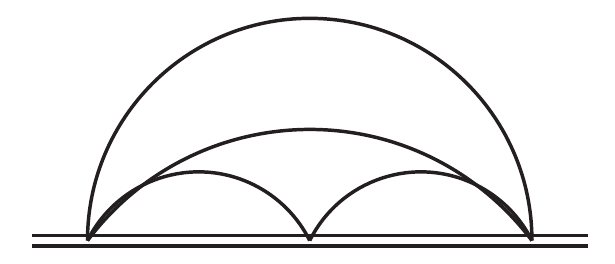} \hspace{2mm}
\includegraphics[scale=0.4]{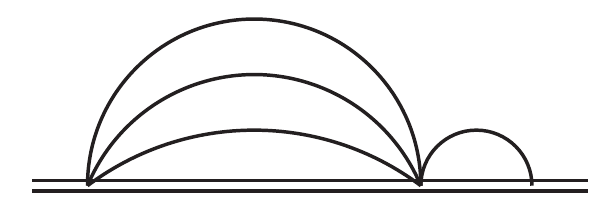} \\[2mm]
\includegraphics[scale=0.4]{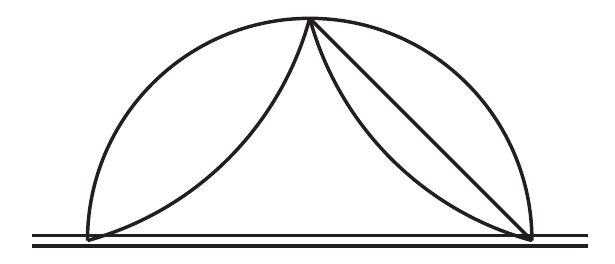} \hspace{2mm}
\includegraphics[scale=0.4]{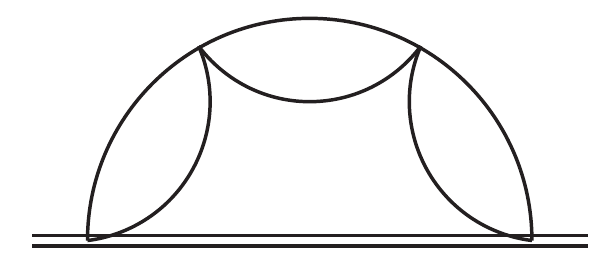} \hspace{2mm}
\includegraphics[scale=0.4]{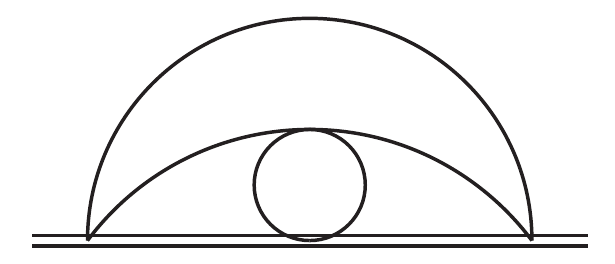} \hspace{2mm}
\includegraphics[scale=0.4]{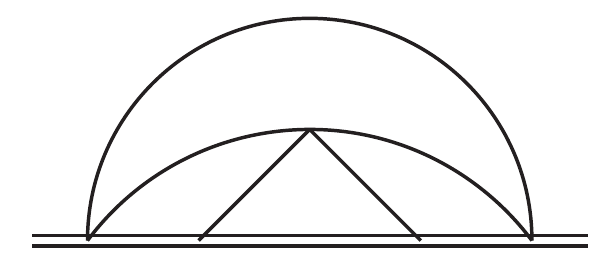} \\[2mm]
\includegraphics[scale=0.4]{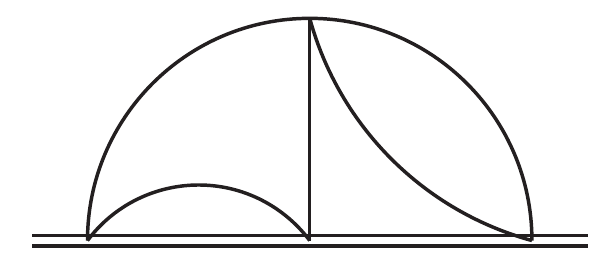} \hspace{2mm}
\includegraphics[scale=0.4]{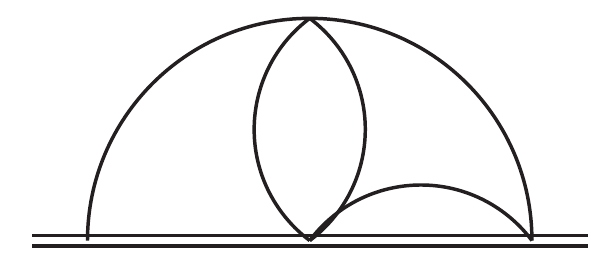} \hspace{2mm}
\includegraphics[scale=0.4]{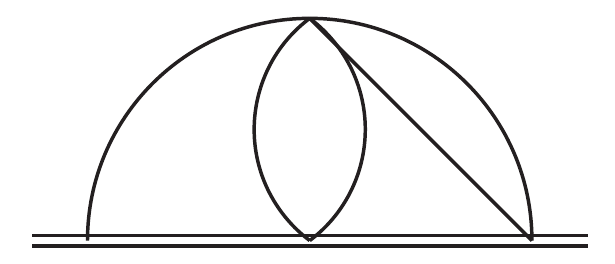} \hspace{2mm}
\includegraphics[scale=0.4]{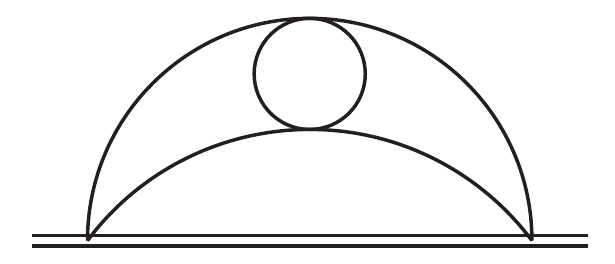} \\[2mm]
\includegraphics[scale=0.4]{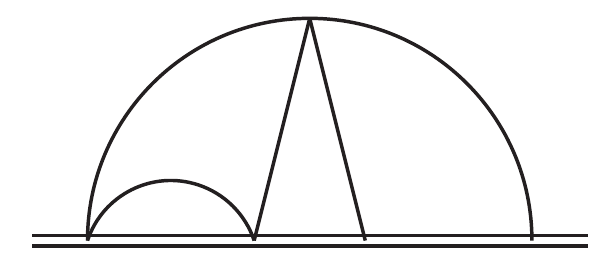} \hspace{2mm}
\includegraphics[scale=0.4]{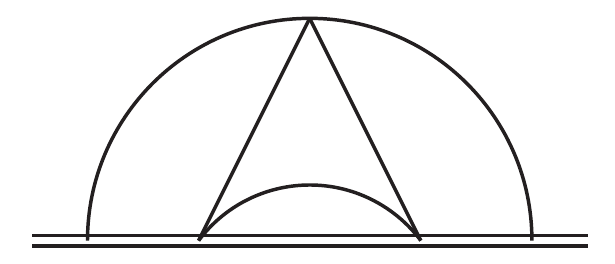} \hspace{2mm}
\includegraphics[scale=0.4]{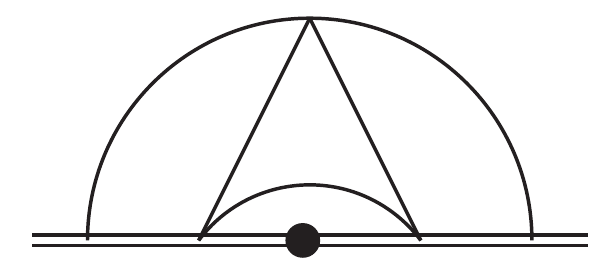} \hspace{2mm}
\includegraphics[scale=0.4]{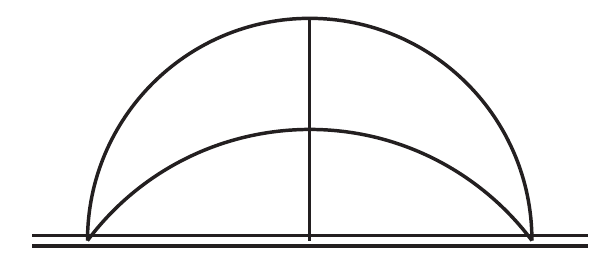} \\[2mm]
\includegraphics[scale=0.4]{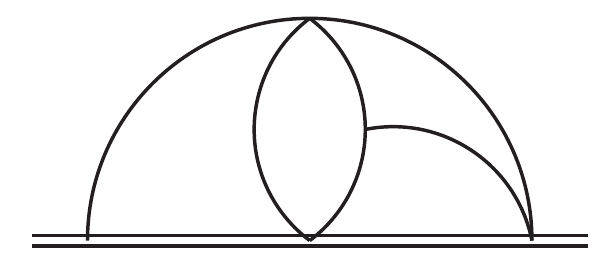} \hspace{2mm}
\includegraphics[scale=0.4]{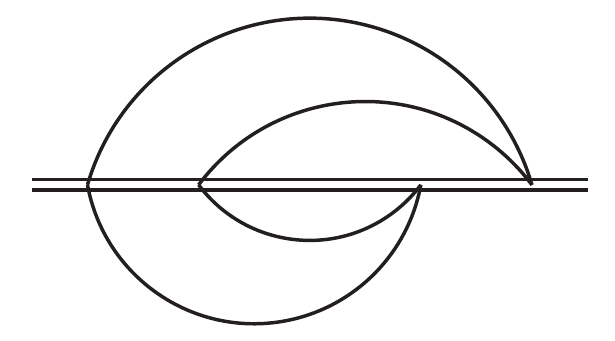} \hspace{2mm}
\includegraphics[scale=0.4]{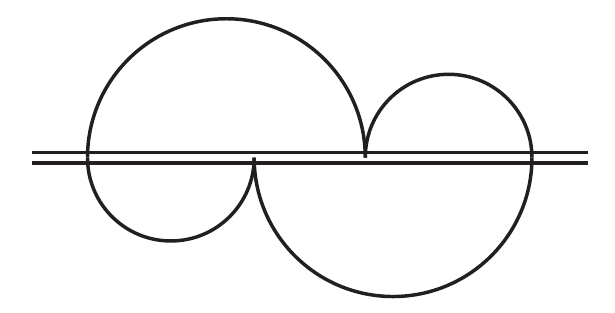} \hspace{2mm}
\includegraphics[scale=0.4]{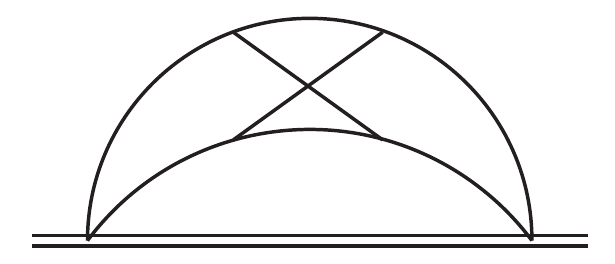} 
\end{center}
\caption{Non-planar master integrals needed for the computation. The double line represents an HQET propagator.}\label{fig:master}
\end{figure}
The last non-planar topology emerges exclusively from diagram (p$_8$) in Figure \ref{fig:pNP} and can thus be discarded in the $N_1=N_2$ case. Additional non-planar master integrals do arise in the master integral reduction of some diagrams, but they eventually do not contribute to the cusp anomalous dimension.

Ultraviolet and infrared divergences are dealt with dimensional regularization ($d=3-2\epsilon$) in the dimensional reduction (DRED) scheme \cite{Siegel:1979wq} and the introduction of a residual energy for the heavy quark.  The final result does not depend on the infrared regulator. 
The $\epsilon$ expansion of the master integrals up to the order required in the computation is given in appendix \ref{app:masters}. 

Putting everything together we get the results for the $\epsilon$ expansions of the single diagrams, which we collect in appendix \ref{app:diagrams}. 
Keeping only the part relevant for the $\theta$-Bremsstrahlung function, the (logarithm of the) 1PI cusp expectation value at $\varphi=0$ at four loops evaluates
\begin{align}\label{eq:Wnonp}
\log W\, \Big|_{ \begin{minipage}{10mm} \scriptsize
$\varphi=0$\\ 
$\theta$-dep
\end{minipage}}
 =& \frac{C_\theta^2 N_1 N_2}{4 k^2 \epsilon }-\frac{N_2 C_{\theta }^4}{48 k^4 \epsilon } \left(N_2 N_1^2 \left(6 C_{\theta }^4+5 \pi ^2-12\right)+\pi ^2 N_2^2 N_1\right.\\&\left.
 +\pi ^2 N_1 \left(1-2 C_{\theta }^4\right)+N_2 \left(-6 C_{\theta }^4-5 \pi ^2+12\right)\right)+{\cal O}\left(k^{-6}\right) + {\cal O}\left(\epsilon^0\right)\nonumber
\end{align}
where $C_\theta = \cos\frac{\theta}{2}$. 

We have several consistency checks on the new, non-planar part.
Since four loops is the first order where sub-leading corrections occur, only simple poles in $\epsilon$ are expected. This is the case in \eqref{eq:Wnonp}, as a result of the non-trivial cancellation of cubic divergences among diagrams (b), (g), (h), (r), (s) and (u) and of double poles originating also from various additional diagrams. Moreover, various graphs involve the one-loop corrected gauge propagator. This contains a non-gauge covariant piece, which is expected not to contribute to physical, gauge invariant expectation values (see e.g. discussions in \cite{Drukker:2008zx,Griguolo:2012iq}). We have indeed verified that such terms drop out of the final result.

From equation \eqref{eq:Wnonp} we can extract $\Gamma_{1/6}(\varphi,\theta)$ and then get the $\theta$-Bremsstrahlung function 
\begin{equation}\label{eq:bremsstrahlungnp}
B_{1/6}^{\theta}(k,N_1,N_2) = \frac{N_1 N_2}{4 k^2}-\frac{\pi ^2 N_2 \left(5 N_1^2 N_2+N_1 N_2^2-3N_1-5 N_2\right)}{24 k^4}+{\cal O}\left(k^{-6}\right)
\end{equation}
for generic ranks of the gauge groups. 
We observe that this result exhibits maximal degree of transcendentality, notwithstanding the mixed transcendentality of the expectation value \eqref{eq:Wnonp}.

\section{Matrix model computation}\label{sec:matrixmodel}

In the planar limit the geometric angle Bremsstrahlung was argued to be obtained as a derivative of a multiply wound 1/6-BPS Wilson loop \cite{Lewkowycz:2013laa}. Moreover, in \cite{Bianchi:2017afp} we proposed that also the $\theta$-Bremsstrahlung function could be obtained this way (with a relative factor of 2 compared to $B_{1/6}^{\varphi}$) and extended the conjecture to different gauge group ranks.
In order to check a possible extension to the color sub-leading case, we need an explicit expansion of the multiply wound 1/6-BPS Wilson loop to four loops.

We computed such a quantity from the matrix model \cite{Kapustin:2009kz,Marino:2009jd,Drukker:2010nc}
\begin{align}
\label{eq:matrix}
Z = &\int \prod_{a=1}^{N_1}d\lambda _{a} \ e^{i\pi k\lambda_{a}^{2}}\prod_{b=1}^{N_2}d\mu_{b} \ e^{-i\pi k\mu_{b}^{2}}\,  \frac{\prod_{a<b}^{N_1}\sinh ^{2}(\pi (\lambda_{a}-\lambda _{b}))\prod_{a<b}^{N_2}\sinh ^{2}(\pi (\mu_{a}-\mu_{b}))}{\prod_{a=1}^{N_1}\prod_{b=1}^{N_2}\cosh ^{2}(\pi (\lambda _{a}-\mu_{b}))}
\end{align}
computing the average
\begin{equation}
\langle W_n \rangle(k,N_1,N_2) = \frac{1}{N_1}\left\langle \sum_{i=1}^{N_1} e^{n\, \lambda_i} \right\rangle
\end{equation}
For the purposes of this paper a pedestrian expansion of the matrix model average at weak coupling is sufficient. This boils down to computing the relevant multi-trace correlators in a Gaussian model (whose explicit expressions can be found in appendix \ref{app:matrix}), after which we obtain the following expression
\begin{align}\label{eq:Wn}
\langle W_n \rangle(k,N_1,N_2) =& 1+\frac{i \pi  n^2 N_1}{k}-\frac{\pi ^2 n^2 \left(n^2 \left(2 N_1^2+1\right)+2 N_1^2-6 N_1 N_2-2\right)}{6 k^2}\nonumber\\&
-\frac{i \pi ^3 n^2}{18 k^3} \left(n^4 \left(N_1^3+2 N_1\right)+n^2 \left(4 N_1^3-12 N_2 N_1^2-4 N_1-6 N_2\right) \right.\nonumber\\&\left.+N_1^3+9 N_1 N_2^2-N_1-6 N_1^2 N_2-3 N_2\right)\nonumber\\&
+\frac{\pi ^4 n^2}{360 k^4} \left(n^6 \left(2 N_1^4+10 N_1^2+3\right)+20 n^4 \left(N_1^4-3 N_2 N_1^3-6 N_2 N_1-1\right) \right.\nonumber\\&+2 n^2 \left(13 N_1^4-75 N_2 N_1^3+5 \left(24 N_2^2-5\right) N_1^2+15 N_2 N_1 +60 N_2^2+12\right)\nonumber\\&\left.-60 N_2 \left(N_2 N_1^2+\left(N_2^2-1\right) N_1-N_2\right)\right) +{\cal O}\left(k^{-5}\right)
\end{align}
some of whose perturbative orders can be checked to agree with available expressions in literature in the planar case and/or for single winding \cite{Drukker:2008zx,Chen:2008bp,Rey:2008bh,Kapustin:2009kz,Marino:2009jd,Drukker:2010nc,Klemm:2012ii,Bianchi:2013zda,Bianchi:2013rma,Griguolo:2013sma,Lewkowycz:2013laa,Bianchi:2016yzj,Bianchi:2016gpg}.
Taking the derivative with respect to the winding number, we derive the following conjectural predictions for the Bremsstrahlung functions
\begin{align}\label{eq:prediction}
B_{1/6}^{\varphi}(k,N_1,N_2) &\underset{\text{conj}}{=} 2\, B_{1/6}^{\theta}(k,N_1,N_2) \underset{\text{conj}}{=} \frac{1}{4\pi^2}\, \partial_n\, |W_n(k,N_1,N_2)|\, \bigg|_{n=1}  \nonumber\\& = \frac{N_1 N_2}{2 k^2}-\frac{\pi ^2 N_2 \left(5 N_1^2 N_2+N_1 N_2^2-3N_1-5 N_2\right)}{12 k^4} +{\cal O}\left(k^{-6}\right)
\end{align}
We observe that albeit the Wilson loop \eqref{eq:Wn} features a plethora of color structures encompassing all possible ones ($N_2^l$ at $l$ loops is in general not possible for this observable by construction), only a subset of them survives in \eqref{eq:prediction}. We comment more on these aspects in the next section.

\section{Conclusions}

The main result of the paper is the agreement between the prediction \eqref{eq:prediction} and the perturbative computation \eqref{eq:bremsstrahlungnp}.
On the one hand this further underpins the conjecture for $B_{1/6}^{\theta}$ put forward in \cite{Bianchi:2017afp}, adding two more data points that verify the relation.
On the other hand the result of this paper hints at its validity including non-planar effects.
In conclusion we propose the conjecture
\begin{align}\label{eq:conjecture}
B_{1/6}^{\varphi}(k,N_1,N_2) &\underset{\text{conj}}{=} 2\, B_{1/6}^{\theta}(k,N_1,N_2) \underset{\text{conj}}{=} \frac{1}{4\pi^2}\, \partial_n\, |W_n(k,N_1,N_2)|\, \bigg|_{n=1} 
\end{align}
holds for all values of the parameters $k$, $N_1$ and $N_2$.

We stress that such an agreement is practically tested against six different data points: the two-loop computation \cite{Bianchi:2014laa} and the five coefficients of the four-loop result, proportional to the different color structures $N_1^3 N_2$ $N_1^2 N_2^2$, $N_1 N_2^3$, $N_1 N_2$ and $N_2^2$.
These are not all the possible color structures, but it is obvious from the perturbative computation that the others vanish trivially. For instance $N_2^l$ at $l$ loops cannot be generated by construction, whereas $N_1^l$ would correspond  to a pure Chern-Simons contribution to the cusp anomalous dimension which we do not expect.
Actually, also the color structure $N_1^3 N_2$ vanishes in the four-loop result \eqref{eq:bremsstrahlungnp}, but non-trivially, as a result of cancellations between diagrams and we counted it as a data point. In fact, as remarked in \cite{Bianchi:2017afp} the localization prediction suggests that the color term $N_1^{l-1}N_2$ at loop $l$ is not present in the Bremsstrahlung function, but we lack an explanation of this phenomenon both at the level of the matrix model prediction and of perturbation theory.

As observed in \cite{Bianchi:2017afp} we stress that the Bremsstrahlung function, including the color sub-leading corrections, is proportional to an overall factor $N_2$. We can interpret this occurrence from Feynman diagrammatics, by recalling that the computation of the $\theta$-Bremsstrahlung function requires the presence of at least one two-point function of bi-scalar insertions. This practically forces the presence of such a color factor. In particular, it forbids the appearance of an otherwise possible term of order $N^0$ in \eqref{eq:bremsstrahlungnp}.
This fact has a rather clear explanation from the Feynman diagram expansion viewpoint, but it lacks an obvious interpretation (at least to us) from the matrix model computation.
In particular this observation seems to pose a set of $l/2+1$ constraints on the perturbative expansion of $\log |\langle W_n \rangle|$ at (even) $l$ loops, stating that the coefficients of certain color structures (those surviving the $N_2\to 0$ limit) have an extremum at $n=1$. These are not particularly powerful constraints in practice, but it would be interesting to understand their origin from the matrix model computation.

While the planar part of this result might be amenable of a complementary derivation using integrability, following results in ${\cal N}=4$ SYM \cite{Correa:2012at,Correa:2012hh,Drukker:2012de,Gromov:2012eu,Gromov:2013qga,Gromov:2015dfa}, non-planar corrections are probably not embraced by this framework (at least according to the current understanding of it). 
Still, it looks that they can be computed exactly using localization results.
Finally, we recall that despite progress on the strong coupling description of Bremsstrahlung functions in ABJM \cite{Correa:2014aga,Aguilera-Damia:2014bqa}, no direct results are available at the moment for the $\theta$-Bremsstrahlung, let alone non-planar effects.
We hope that our results could possibly shed more light on the strong coupling picture of the $\theta$-Bremsstrahlung function.

\acknowledgments

This work has been supported in part by Italian Ministero dell'Istruzione, Universit\`a e Ricerca (MIUR) and Istituto
Nazionale di Fisica Nucleare (INFN) through the ``Gauge Theories, Strings, Supergravity'' (GSS) research project.

\vfill
\newpage

\appendix

\section{Master integrals definitions and expansions}\label{app:masters}

The (Euclidean) HQET planar integrals at four loops were defined in \cite{Bianchi:2017afp} by the following products of propagators ($d=3-2\epsilon$)
\begin{align}\label{eq:masterintegrals}
G_{a_1,\dots,a_{14}} \equiv \int \frac{d^dk_1\, d^dk_2\, d^dk_3\, d^dk_4}{(2\pi)^{4d}}\, \prod_{i=1}^{14}\frac{1}{P_i^{a_i}}
\end{align}
where the explicit propagators read
\begin{align}
& P_1 = (2k_1\cdot \tilde v+1) ,\quad P_2 = (2k_2\cdot \tilde v+1) ,\quad P_3 = (2k_3\cdot \tilde v+1) ,\quad P_4 = (2k_4\cdot \tilde v+1) \nonumber\\&
P_5 = k_1^2 ,\qquad P_6 = k_2^2 ,\qquad P_7 = k_3^2 ,\qquad P_8 = k_4^2 \nonumber\\&
P_9 = (k_1-k_2)^2 ,\qquad P_{10} = (k_1-k_3)^2 ,\qquad P_{11} = (k_1-k_4)^2  \nonumber\\&
P_{12} = (k_2-k_3)^2 ,\qquad P_{13} = (k_2-k_4)^2 ,\qquad P_{14} = (k_3-k_4)^2
\end{align}
and $\tilde v^2=-1$. Non-planar master integrals can be defined from \eqref{eq:masterintegrals} with the following two sets of propagators
\begin{align}
& P_1^{\text{NP}_1} = (2k_1\cdot \tilde v+1) ,\, P_2^{\text{NP}_1} = (2k_2\cdot \tilde v+1) ,\, P_3^{\text{NP}_1} = (2k_3\cdot \tilde v+1) ,\, P_4^{\text{NP}_1} = (2k_4\cdot \tilde v+1) \nonumber\\&
P_5^{\text{NP}_1} = k_2^2 ,\qquad P_6^{\text{NP}_1} = k_3^2 ,\qquad P_7 = k_4^2 ,\qquad P_8^{\text{NP}_1} = (k_1-k_2)^2 \nonumber\\&
P_{9}^{\text{NP}_1} = (k_1-k_3)^2 ,\qquad P_{10}^{\text{NP}_1} = (k_1-k_4)^2 ,\qquad P_{11}^{\text{NP}_1} = (k_2-k_4)^2  \nonumber\\&
P_{12}^{\text{NP}_1} = (k_3-k_4)^2,\qquad P_{13}^{\text{NP}_1} = (k_1-k_3-k_4)^2,\qquad P_{14}^{\text{NP}_1} = (k_2-k_3-k_4)^2
\end{align}
and
\begin{align}
& P_1^{\text{NP}_2} = (2k_1\cdot \tilde v+1) ,\, P_2^{\text{NP}_2} = (2k_2\cdot \tilde v+1) ,\, P_3^{\text{NP}_2} = (2k_3\cdot \tilde v+1) ,\, P_4^{\text{NP}_2} = (2k_4\cdot \tilde v+1) \nonumber\\&
P_5^{\text{NP}_2} = k_1^2 ,\qquad P_6^{\text{NP}_2} = k_2^2 ,\qquad P_7^{\text{NP}_2} = k_3^2 ,\qquad P_8^{\text{NP}_2} = k_4^2 \nonumber\\&
P_9^{\text{NP}_2} = (k_1-k_2)^2 ,\qquad P_{10}^{\text{NP}_2} = (k_1-k_3)^2 ,\qquad P_{11}^{\text{NP}_2} = (k_1-k_4)^2  \nonumber\\&
P_{12}^{\text{NP}_2} = (k_2-k_3)^2 ,\qquad P_{13}^{\text{NP}_2} = (k_2-k_4)^2 ,\qquad P_{14}^{\text{NP}_2} = (k_1+k_2-k_3-k_4)^2
\end{align}
The required topologies fall within these categories taking subsets of their indices to be non-negative.
We used these definitions for reductions with LiteRed and FIRE.

The computation presented in this paper entails the expansion of the 21 master integrals of \cite{Bianchi:2017afp} to certain orders in $\epsilon$.
From reduction of the relevant non-planar diagrams it turns out that only three non-planar master integrals are needed for the computation (other non-planar integrals arise in tensor reduction of some diagrams, but eventually do not contribute to the four-loop Bremsstrahlung function).
These are 
\begin{align}
G^{\text{NP}_1}(1, 1, 1, 0, 0, 0, 1, 0, 0, 0, 1, 0, 1, 1) &= \raisebox{-5mm}{\includegraphics[scale=0.4]{MINP1}} \\
G^{\text{NP}_1}(1, 1, 1, 0, 0, 0, 1, 0, 0, 1, 1, 0, 0, 1) &= \raisebox{-5mm}{\includegraphics[scale=0.4]{MINP2}} \\
G^{\text{NP}_2}(1,0,0,0,0,1,1,1,0,1,1,1,1,1) & = \raisebox{-4mm}{\includegraphics[scale=0.4]{MINP3}}
\end{align}
Their evaluation up to the required orders in $\epsilon$ can be attained straightforwardly by standard techniques.

We start recalling the expressions for the basic bubble integrals
\begin{align}
\begin{minipage}{3cm}\vspace{-0.45cm}
\includegraphics[scale=0.35]{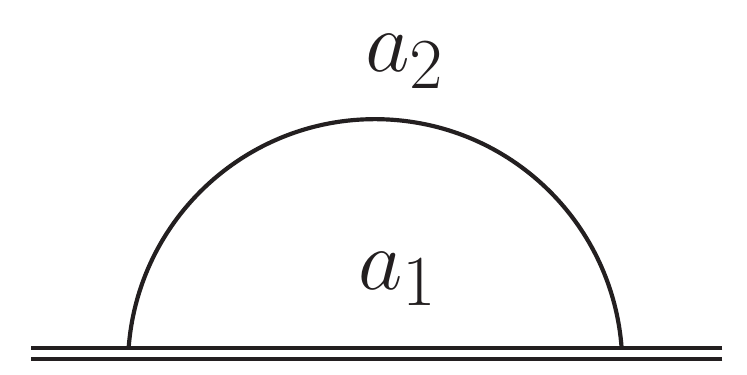}
\end{minipage} &= I(a_1,a_2) = G_{a_1,a_2}  = \frac{1}{(4 \pi)^{d/2}}\frac{\Gamma(a_1+ 2 a_2-d) \Gamma(d/2-a_2)}{\Gamma(a_1)\Gamma(a_2)}
\\[4mm]
\begin{minipage}{3cm}\vspace{-0.45cm}
\includegraphics[scale=0.525]{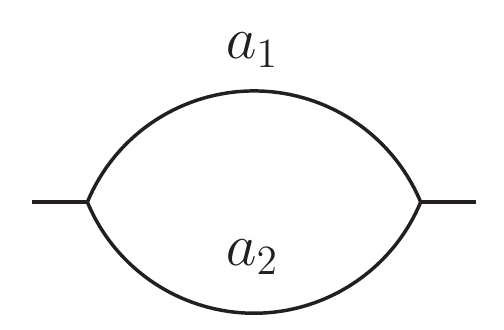}
\end{minipage} &= B_{a_1,a_2}  = \frac{1}{(4 \pi)^{d/2}}\frac{\Gamma(d/2-a_1)\Gamma(d/2-a_2)\Gamma(a_1 + a_2 - d/2)}{\Gamma(a_1)\Gamma(a_2)\Gamma(d - a_1 - a_2 )}
\end{align}
where in the following we drop the $4\pi$ normalization and $\gamma_E$ factors.

Integral $G^{\text{NP}_1}(1, 1, 1, 0, 0, 0, 1, 0, 0, 1, 1, 0, 0, 1)$ is the easiest to evaluate. 
After integrating the two bubble integrals we obtain
\begin{equation}
\raisebox{-4mm}{\includegraphics[scale=0.4]{MINP2}} = I^2(1,1)\, J(2\epsilon, 2\epsilon,1,1,1)
\end{equation}
where the function $J$ corresponds to the two-loop integral with generic indices, and which was computed in \cite{Grozin:2003ak}
\begin{align}
& J(n_1,n_2,n_3,n_4,n_5) = \raisebox{-10mm}{\includegraphics[scale=0.5]{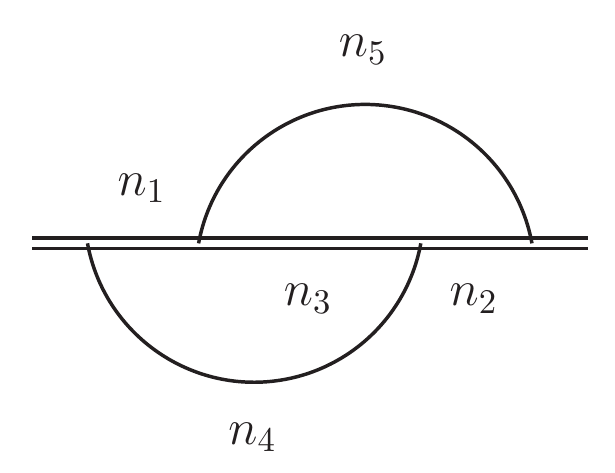}} \nonumber\\
=& \frac{\Gamma \left(\frac{d}{2}-n_4\right) \Gamma \left(\frac{d}{2}-n_5\right) \Gamma \left(-d+n_1+n_3+2 n_4\right) \Gamma \left(-2 d+n_1+n_2+n_3+2 n_4+2 n_5\right)}{\Gamma \left(n_1+n_3\right) \Gamma \left(n_4\right) \Gamma \left(n_5\right) \Gamma \left(-d+n_1+n_2+n_3+2 n_4\right)}\nonumber\\&
\times {}_3F_2\left(\left.\begin{array}{ccc}
& n_1, -d+n_1+n_3+2 n_4, d-2 n_5 & \\
 & n_1+n_3,-d+n_1+n_2+n_3+2 n_4 & 
\end{array}
\right|1\right)
\end{align}
Integral $G^{\text{NP}_1}(1, 1, 1, 0, 0, 0, 1, 0, 0, 0, 1, 0, 1, 1)$ can be dealt with first integrating the innermost bubble integral and then Mellin-Barnes decomposing the resulting expression. The result consists of a one-fold integral, whose expansion coefficients can be directly reduced to numbers, up to the required order in $\epsilon$.

Finally, integral $G^{\text{NP}_2}(1,0,0,0,0,1,1,1,0,1,1,1,1,1)$ arises from the three-loop non-planar correction to the scalar two-point function
\begin{equation}
\raisebox{-4mm}{\includegraphics[scale=0.4]{MINP3}} = I(7/2+3\epsilon,1)\, \raisebox{-4mm}{\includegraphics[scale=0.4]{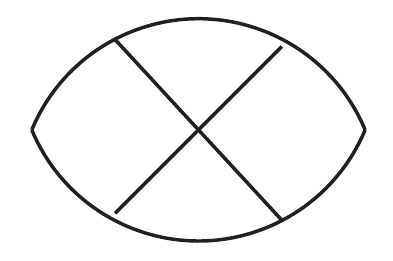}}
\end{equation}
The three-loop non-planar integral can be computed using the GPXT technique \cite{Chetyrkin:1980pr}. 
This turns the integral into a three-fold series of 30 pieces arising from integration of the radial part.
As for the four-dimensional case an overall factor of $\epsilon$ is present which (up to the required order) allows to retain the pole part of the sum only. This simplifies the evaluation substantially, reducing it to a set of at most one-fold sums.
At a difference with four dimensions intermediate double poles are generated in intermediate steps, which cancel in the final answer.
The integral thus evaluates \cite{Young:2014lka}
\begin{equation}
\raisebox{-4mm}{\includegraphics[scale=0.4]{2ptNP}} = \frac{8}{3} \pi ^{5/2} \left(2 \pi ^2-39\right)+O\left(\epsilon ^1\right)
\end{equation}

Altogether, the expansions of the master integrals (including the planar ones already evaluated in \cite{Bianchi:2017afp}) read
\begin{align}
\raisebox{-4mm}{\includegraphics[scale=0.4]{MI1}} &= -\frac{\pi ^2}{48 \epsilon }-\frac{1}{36} \pi ^2 (11+6 \log 2) \nonumber\\& -\frac{1}{432} \left(\pi ^2 \left(57 \pi ^2+8 (170+6 \log 2 (22+6 \log 2))\right)\right) \epsilon +{\cal O}\left(\epsilon ^2\right) \\
\raisebox{-1.5mm}{\includegraphics[scale=0.4]{MI2}} &= \frac{\pi ^2}{16 \epsilon ^2}+\frac{\pi ^2 (2+2 \log 2)}{4 \epsilon }\nonumber\\& +\frac{1}{48} \pi ^2 \left(11 \pi ^2+24 (6+2 \log 2 (4+2 \log 2))\right)+{\cal O}\left(\epsilon ^1\right) \\
\raisebox{-1.5mm}{\includegraphics[scale=0.4]{MI4}} &= -\frac{\pi ^2}{12 \epsilon ^3}-\frac{\pi ^2 (3+4 \log 2)}{6 \epsilon ^2} \nonumber\\& -\frac{\pi ^2 \left(11 \pi ^2+12 (9+4 \log 2 (3+2 \log 2))\right)}{36 \epsilon }
+{\cal O}\left(\epsilon ^0\right) \\
\raisebox{-4mm}{\includegraphics[scale=0.4]{MI21}} &= -\frac{3 \pi ^2}{32 \epsilon ^3}-\frac{3 \left(\pi ^2 (1+\log 2)\right)}{4 \epsilon ^2} \nonumber\\& -\frac{\pi ^2 \left(13 \pi ^2+96 \left(2+2 \log 2+\log ^2 2\right)\right)}{32 \epsilon }
+{\cal O}\left(\epsilon ^0\right) \\
\raisebox{-4mm}{\includegraphics[scale=0.4]{MI6}} &= \frac{3 \pi ^2}{64 \epsilon ^2}+\frac{3 \pi ^2 (5+4 \log 2)}{32 \epsilon } \nonumber\\& +\frac{3}{64} \pi ^2 \left(5 \pi ^2+8 (10+2 \log 2 (5+2 \log 2))\right)+{\cal O}\left(\epsilon ^1\right) \\
\raisebox{-4mm}{\includegraphics[scale=0.4]{MI5}} &= -\frac{\pi ^4}{8 \epsilon }-\frac{1}{2} \pi ^4 (2+3 \log 2)+{\cal O}\left(\epsilon ^1\right) \\
\raisebox{-4mm}{\includegraphics[scale=0.4]{MI7}} &= \frac{\pi ^2}{16 \epsilon ^2}+\frac{\pi ^2 (3+2 \log 2)}{4 \epsilon } \nonumber\\& +\frac{1}{48} \pi ^2 \left(17 \pi ^2+24 (14+2 \log 2 (6+2 \log 2))\right)+{\cal O}\left(\epsilon ^1\right) \\
\raisebox{-4mm}{\includegraphics[scale=0.4]{MI8}} &= \frac{\pi ^2}{24 \epsilon ^2}+\frac{\pi ^2 (9+8 \log 2)}{24 \epsilon } \nonumber\\& +\frac{1}{72} \pi ^2 \left(189+13 \pi ^2+216 \log 2+96 \log ^2 2\right)+{\cal O}\left(\epsilon ^1\right) \\
\raisebox{-4mm}{\includegraphics[scale=0.4]{MI9}} &= -\frac{\pi ^2}{8 \epsilon ^2}-\frac{\pi ^2 (7+4 \log 2)}{4 \epsilon } \nonumber\\& -\frac{1}{24} \pi ^2 \left(11 \pi ^2+12 (37+4 \log 2 (7+2 \log 2))\right)+{\cal O}\left(\epsilon ^1\right) \\
\raisebox{-4mm}{\includegraphics[scale=0.4]{MI20}} &= -\frac{\pi ^4}{2 \epsilon }-3 \left(\pi ^4 (-1+2 \log 2)\right)+{\cal O}\left(\epsilon ^1\right) \\
\raisebox{-4mm}{\includegraphics[scale=0.4]{MI11}} &= \frac{\pi ^4}{48 \epsilon ^2}+\frac{\frac{1}{2} \pi ^4 \log 2-\frac{7 \pi ^2 \zeta (3)}{8}}{\epsilon }+{\cal O}\left(\epsilon ^0\right) \\
\raisebox{-4mm}{\includegraphics[scale=0.4]{MI15}} &= \frac{2 \pi ^4}{3}+{\cal O}\left(\epsilon ^1\right) \\
\raisebox{-4mm}{\includegraphics[scale=0.4]{MI3}} &= \frac{\pi ^4}{16 \epsilon ^2}+\frac{\pi ^4 \log 2-\frac{7 \pi ^2 \zeta (3)}{8}}{\epsilon }+{\cal O}\left(\epsilon ^0\right) \\
\raisebox{-4mm}{\includegraphics[scale=0.4]{MI12}} &= \frac{\pi ^2}{24 \epsilon ^3}+\frac{\pi ^2 (5+4 \log 2)}{12 \epsilon ^2} \nonumber\\& +\frac{\pi ^2 \left(13 \pi ^2+12 (25+4 \log 2 (5+2 \log 2))\right)}{72 \epsilon }+{\cal O}\left(\epsilon ^0\right) \\
\raisebox{-4mm}{\includegraphics[scale=0.4]{MI13}} &= {\cal O}\left(\frac{1}{\epsilon }\right) \\
\raisebox{-4mm}{\includegraphics[scale=0.4]{MI10}} &= -\frac{\pi ^4}{2 \epsilon }+{\cal O}\left(\epsilon ^0\right) \\
\raisebox{-4mm}{\includegraphics[scale=0.4]{MI16}} &= \frac{2 \pi ^4}{3 \epsilon }+{\cal O}\left(\epsilon ^0\right) \\
\raisebox{-4mm}{\includegraphics[scale=0.4]{MI17}} &= {\cal O}\left(\frac{1}{\epsilon }\right) \\
\raisebox{-4mm}{\includegraphics[scale=0.4]{MI18}} &= \frac{2 \pi ^4}{3}+{\cal O}\left(\epsilon ^1\right) \\
\raisebox{-4mm}{\includegraphics[scale=0.4]{MI19}} &= \frac{\pi ^2}{2 \epsilon ^2}+\frac{4 \pi ^2 \log 2}{\epsilon }+\frac{1}{6} \pi ^2 \left(-84+7 \pi ^2+96 \log ^2 2\right)+{\cal O}\left(\epsilon ^1\right) \label{4Lmast1}\\
\raisebox{-4mm}{\includegraphics[scale=0.4]{MI14}} &= \frac{\pi ^2}{2 \epsilon ^2}+\frac{4 \pi ^2 \log 2}{\epsilon }+\frac{1}{6} \pi ^2 \left(-180+23 \pi ^2+96 \log ^2 2\right)+{\cal O}\left(\epsilon ^1\right)\\
\raisebox{-5.25mm}{\includegraphics[scale=0.4]{MINP1}} &= -\frac{\pi ^2}{32 \epsilon ^3}-\frac{\pi ^2 (1+\log 2)}{4 \epsilon ^2}-\frac{\pi ^2 \left(13 \pi ^2+96 \left(2+\log ^2 2+\log 4\right)\right)}{96 \epsilon }+{\cal O}\left(\epsilon ^0\right) \\
\raisebox{-5.25mm}{\includegraphics[scale=0.4]{MINP2}} &= -\frac{\pi ^2}{16 \epsilon ^3}-\frac{\pi ^2 (1+\log 2)}{2 \epsilon ^2}-\frac{\pi ^2 \left(5 \pi ^2+32 \left(2+\log ^2 2+2 \log 2\right)\right)}{16 \epsilon }+{\cal O}\left(\epsilon ^0\right) \\
\raisebox{-4mm}{\includegraphics[scale=0.4]{MINP3}} &= \frac{1}{\epsilon }\left(\frac{832 \pi ^2}{5}-\frac{128 \pi ^4}{15}\right)+{\cal O}\left(\epsilon ^0\right)
  \label{4Lmast2}
\end{align}
where an overall factor $e^{-4 \gamma_E \epsilon}/(4\pi)^{2d}$ is omitted.

\section{Results for the four-loop diagrams}\label{app:diagrams}

In this appendix we list the results for the diagrams of Figure \ref{fig:4loops} and  \ref{fig:4loopsNP}, including the color sub-leading factors. A common factor $\left(\frac{e^{-4\epsilon\gamma_E}}{k(4\pi)^{d/2}}\right)^4$ is understood.
The planar topologies of Figure \ref{fig:4loops}  yield
\begin{align}
(a)= & \frac{8 \pi ^2 N_1^2 N_2^2 C_{\theta }^2 \left(C_{\theta }^2-2\right)}{\epsilon ^2}+\frac{32 \pi ^2 N_1^2 N_2^2 C_{\theta }^2 \left((2 \log 2-1) C_{\theta }^2-4 \log 2\right)}{\epsilon }+O\left(\epsilon ^0\right) \\
(b)= & \frac{4 \pi ^2 N_1 \left(N_1^2+1\right) N_2 C_{\theta }^2}{\epsilon ^3}+\frac{32 \pi ^2 N_1 \left(N_1^2+1\right) N_2 \log 2 C_{\theta }^2}{\epsilon ^2}\nonumber\\&+\frac{4 N_1 \left(N_1^2+1\right) N_2 \left(13 \pi ^4+96 \pi ^2 \log ^2(2)\right) C_{\theta }^2}{3 \epsilon }+O\left(\epsilon ^0\right) \\
(c)= & -\frac{16 \left(\pi ^2 N_1^2 N_2^2 C_{\theta }^2\right)}{\epsilon ^2}+\frac{16 \pi ^2 N_1^2 N_2^2 (7-8 \log 2) C_{\theta }^2}{\epsilon }+O\left(\epsilon ^0\right) \\
(d)= & \frac{16 \pi ^2 N_1^2 N_2^2 C_{\theta }^2}{\epsilon ^2}+\frac{16 \pi ^2 N_1^2 N_2^2 (1+8 \log 2) C_{\theta }^2}{\epsilon }+O\left(\epsilon ^0\right) \\
(e)= & \frac{128 \pi ^2 \left(N_1^2-1\right) N_2^2 C_{\theta }^2}{\epsilon }+O\left(\epsilon ^0\right) \\
(f)= & -\frac{32 \left(\pi ^2 \left(\pi ^2-4\right) \left(N_1^2-1\right) N_2^2 C_{\theta }^2\right)}{\epsilon }+O\left(\epsilon ^0\right) \\
(g)= & -\frac{2 \left(\pi ^2 N_1 \left(N_1^2+3\right) N_2 C_{\theta }^2\right)}{\epsilon ^3}-\frac{2 \left(\pi ^2 N_1 \left(N_1^2+3\right) N_2 (8 \log 2-1) C_{\theta }^2\right)}{\epsilon ^2}\nonumber\\&-\frac{4 \left(\pi ^2 N_1 \left(N_1^2+3\right) N_2 \left(-9+7 \pi ^2+12 \log 2 (4 \log 2-1)\right) C_{\theta }^2\right)}{3 \epsilon }+O\left(\epsilon ^0\right) \\
(h)= & -\frac{2 \left(\pi ^2 N_1 \left(N_1^2+3\right) N_2 C_{\theta }^2\right)}{\epsilon ^3}-\frac{2 \left(\pi ^2 N_1 \left(N_1^2+3\right) N_2 (8 \log 2-1) C_{\theta }^2\right)}{\epsilon ^2}\nonumber\\&-\frac{2 \left(\pi ^2 N_1 \left(N_1^2+3\right) N_2 \left(17 \pi ^2+6 (4 \log 2 (4 \log 2-1)-3)\right) C_{\theta }^2\right)}{3 \epsilon }+O\left(\epsilon ^0\right) \\
(i)= & -\frac{2\pi ^4 N_1 N_2\left(N_1^2+3\right) C_{\theta }^2}{3 \epsilon }+O\left(\epsilon ^0\right) \\
(j)= & \frac{8 \pi ^4 (N_1^2 N_2^2-2N_1N_2+N_2^2) C_{\theta }^2}{3 \epsilon }+O\left(\epsilon ^0\right) \\
(k)= & -\frac{8 \pi ^4 (N_1^2 N_2^2-2N_1N_2+N_2^2) C_{\theta }^2}{3 \epsilon }+O\left(\epsilon ^0\right) \\
(l)= & -\frac{4 \pi ^4 (N_1^2-1)N_1 N_2 C_{\theta }^2}{3 \epsilon }+O\left(\epsilon ^0\right) \\
(m)= & \frac{4 \pi ^4 (N_1^2-1)N_1 N_2 C_{\theta }^2}{3 \epsilon }+O\left(\epsilon ^0\right) \\
(n)= & -\frac{8 \pi ^4 (N_1^2 N_2^2-2N_1N_2+N_2^2) C_{\theta }^2}{3 \epsilon }+O\left(\epsilon ^0\right) \\
(o)= & \frac{8 \pi ^4 (N_1^2 N_2^2-2N_1N_2+N_2^2) C_{\theta }^2}{3 \epsilon }+O\left(\epsilon ^0\right) \\
(p) = & -\frac{4 \left(\pi ^2 \left(N_1^2-1\right) N_2 \left(N_1+4 N_2\right) C_{\theta }^2\right)}{\epsilon ^2}\\& +\frac{4 \pi ^2 N_2 C_{\theta }^2}{3 \epsilon } \left(3 N_1^3 \left(-6+\pi ^2-8 \log 2\right)+4 N_2 N_1^2 \left(\pi ^2-6 (7+4 \log 2)\right)\right.\nonumber\\&\left.+N_1 \left(\pi ^2 \left(1-4 N_2^2\right)+6 (3+4 \log 2)\right)-4 N_2 \left(\pi ^2-6 (7+4 \log 2)\right)\right)+O\left(\epsilon ^0\right)\nonumber
\end{align}
We observe that diagrams $(l)$-$(o)$ cancel pairwise. The non-planar diagrams of Figure \ref{fig:4loopsNP} read
\begin{align}
(q)= & N_2^2 C_{\theta }^2\left(\frac{16 \pi ^2 }{\epsilon ^2}+\frac{32 \pi ^2  \left(C_{\theta }^2+4 \log 2\right)}{\epsilon }\right)+O\left(\epsilon ^0\right) \\
(r)= & N_2^2 C_{\theta }^2\left(-\frac{32 \left(\pi ^2 \right)}{\epsilon ^2}-\frac{32 \left(\pi ^2  (1+8 \log 2) \right)}{\epsilon }\right)+O\left(\epsilon ^0\right) \\
(s)= & N_1 N_2 C_{\theta }^2\left(\frac{4 \pi ^2 }{\epsilon ^3}+\frac{32 \pi ^2 \log 2}{\epsilon ^2}+\frac{4 \left(\pi ^4 \left(8 C_{\theta }^2+3\right)+96 \pi ^2 \log ^2 2\right)}{3 \epsilon }\right)+O\left(\epsilon ^0\right) \\
(t)= & N_1 N_2 C_{\theta }^2\left(\frac{4 \pi ^2 }{\epsilon ^3}+\frac{32 \pi ^2 \log 2}{\epsilon ^2}+\frac{4 \pi ^2 \left(19 \pi ^2+96 \log ^2 2\right)}{3 \epsilon }\right)+O\left(\epsilon ^0\right) \\
(u)= & N_1 N_2 C_{\theta }^2\left(-\frac{16 \left(\pi ^2\right)}{\epsilon ^2}+\frac{32 \pi ^2 \left(\pi ^2-3 (3+4 \log 2)\right)}{3 \epsilon }\right)+O\left(\epsilon ^0\right) 
\end{align}

\section{Two-loop scalar propagator corrections}\label{app:prop}

As a by-product of this computation we present here the two-loop corrections to the scalar self-energy, including color sub-leading corrections.
We recall that the planar part was computed in \cite{Minahan:2009wg}.
Sub-leading corrections arise  from different contractions of the planar topologies of \eqref{eq:2loopscalar}
\begin{align}\label{eq:2loopscalar}
\raisebox{-2.5mm}{\includegraphics[scale=0.3]{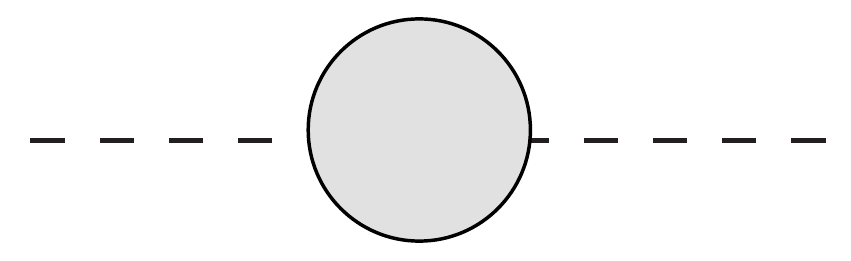}}\, =\, & \raisebox{-2.5mm}{\includegraphics[scale=0.3]{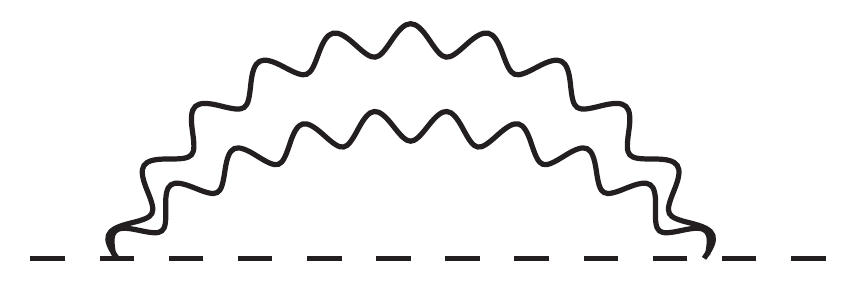}}\, +\, \raisebox{-3.5mm}{\includegraphics[scale=0.3]{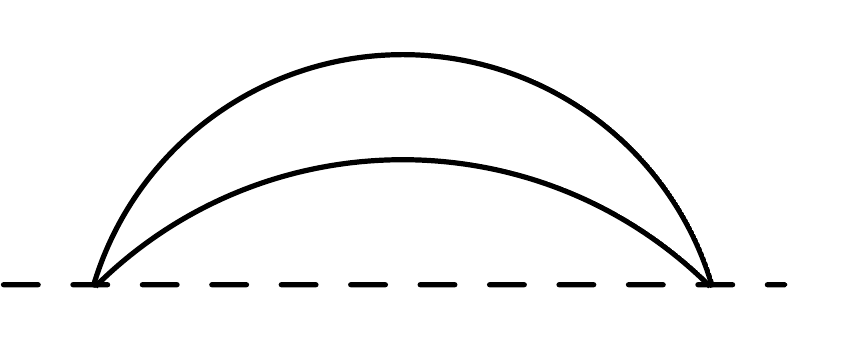}}\, +\, \raisebox{-4mm}{\includegraphics[scale=0.3]{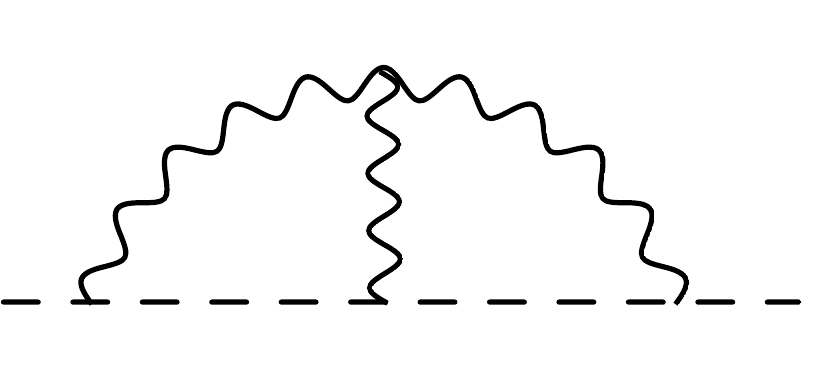}} \nonumber\\&
\raisebox{-6.5mm}{\includegraphics[scale=0.3]{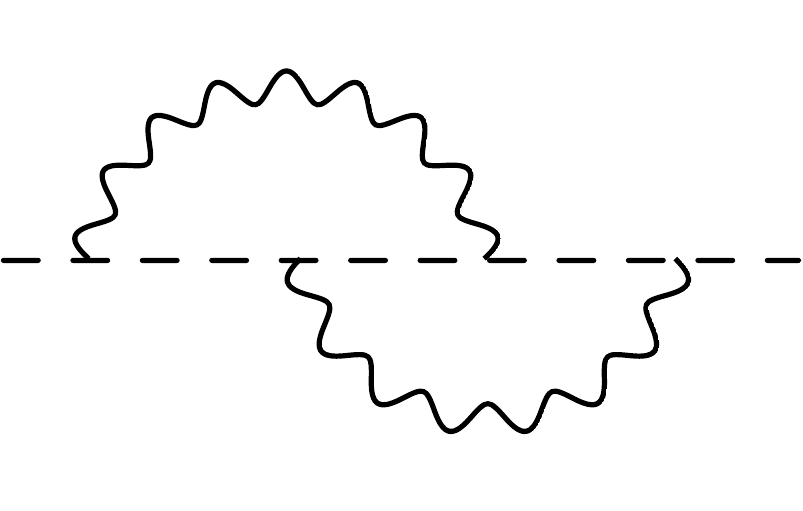}}\, +\, \raisebox{-5mm}{\includegraphics[scale=0.3]{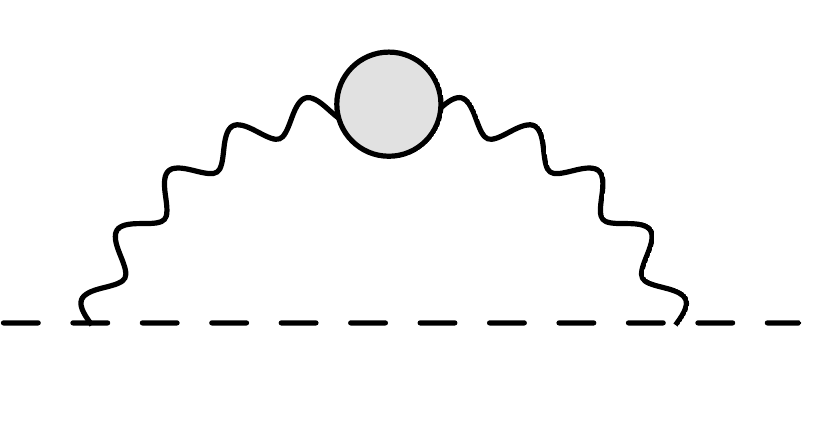}}
\end{align}

\begin{align}
&\raisebox{-2.5mm}{\includegraphics[scale=0.3]{scalarNP1}}\, =\,  N_1 N_2 \left(N_1^2-4 N_2 N_1+N_2^2+2\right)\left(\frac{\pi}{3 \epsilon }+2 \pi +O\left(\epsilon ^1\right)\right) \\
&\raisebox{-3.5mm}{\includegraphics[scale=0.3]{scalarNP5}}\,=\,  N_1 N_2 \left(N_1 N_2-1\right)\left(-\frac{56 \pi}{3 \epsilon }-112 \pi +O\left(\epsilon ^1\right)\right) \\
&\raisebox{-4mm}{\includegraphics[scale=0.3]{scalarNP4}}\, =\, N_1 N_2 \left(N_1^2+N_2^2-2\right)\left(-\frac{4\pi}{3 \epsilon }+\pi  \left(\pi ^2-8\right) +O\left(\epsilon ^1\right)\right) \\
&\raisebox{-6.5mm}{\includegraphics[scale=0.3]{scalarNP2}}\,= \, N_1 N_2 \left(N_1 N_2-1\right)\left(-\frac{16 \pi}{3 \epsilon }+4 \pi  \left(\pi ^2-8\right)+O\left(\epsilon ^1\right)\right) \\
&\raisebox{-5mm}{\includegraphics[scale=0.3]{scalarNP3}}\,=\,  N_1 N_2 \left(N_1 N_2-1\right)\left(\frac{64 \pi}{3 \epsilon }+64 \pi+O\left(\epsilon ^1\right) \right)
\end{align}
The corresponding contributions to diagram (p$_1$) are obtained by multiplying these by $8\, B(1+2\epsilon,1)\, I(2,1/2+3\epsilon)$, where a factor of 2 stems from the two scalar propagators, a factor 4 comes from the normalization of HQET integrals and the indices of the bubble integrals are fixed by dimensional analysis.

\section{Scalar bubble corrections}\label{app:vertex}

\begin{figure}[h]
\centering
\includegraphics[scale=0.3]{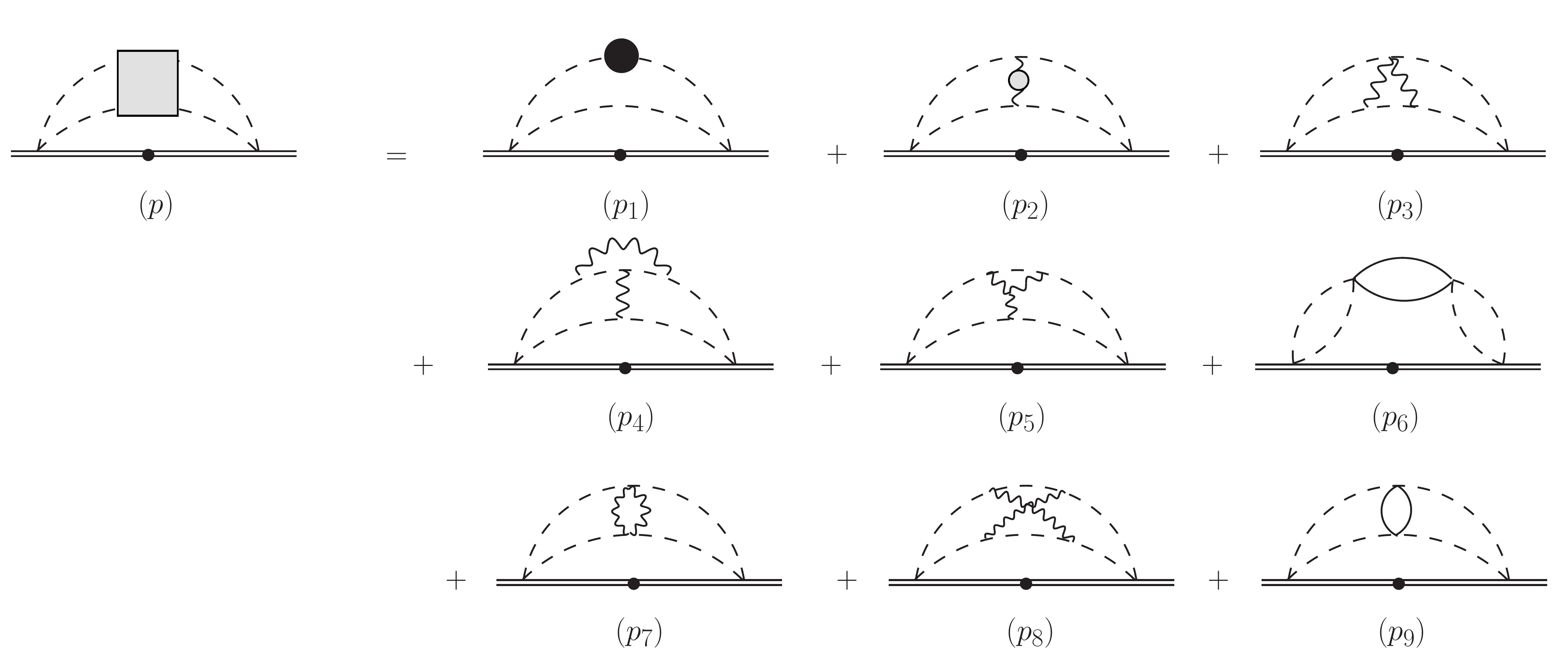}
\caption{Scalar bubble corrections} \label{diag_p}
\end{figure}

Diagram (p) of Figure \ref{fig:4loops} comprises the corrections to the scalar bilinear two-point function. Its non-vanishing  contributions (some possible contractions generate for instance $\Tr M_{1,2} = 0$), including color sub-leading ones are listed in Figure \ref{diag_p}.
In addition, diagram $(p_1)$ involves the 2-loop correction to the scalar propagator, which we detailed in the previous appendix. 
Altogether, the various contributions from diagram (p) to the cusp expectation value read
\begin{align}
(p_1)= & -\frac{4 \left(\pi ^2 N_1 N_2 \left(N_1^2+4 N_2 N_1+N_2^2-6\right) C_{\theta }^2\right)}{\epsilon ^2}\nonumber\\&
+\frac{4 \pi ^2 N_1 N_2 C_{\theta }^2}{\epsilon } \left((N_1^2+N_2^2) \left(-6+\pi ^2-8 \log 2\right)+4 N_2 N_1 \left(-22+\pi ^2-8 \log 2\right)\right.\nonumber\\&\left.-6 \pi ^2+100+48 \log 2\right)+O\left(\epsilon ^0\right) \\
(p_2)= & -\frac{16 \left(\pi ^2 \left(\pi ^2-12\right) N_2 \left(N_2+N_1 \left(N_1 N_2-2\right)\right) C_{\theta }^2\right)}{\epsilon }+O\left(\epsilon ^0\right) \\
(p_3)= & -\frac{4 \left(\pi ^2 \left(\pi ^2-12\right) N_2 \left(N_1 \left(N_2^2+3\right)-4 N_2\right) C_{\theta }^2\right)}{\epsilon }+O\left(\epsilon ^0\right) \\
(p_4)= & \frac{16 \pi ^2 \left(\pi ^2-12\right) N_2 \left(N_2+N_1 \left(N_1 N_2-2\right)\right) C_{\theta }^2}{3 \epsilon }+O\left(\epsilon ^0\right) \\
(p_5)= & \frac{8 \pi ^2 \left(\pi ^2-12\right) N_1 N_2 \left(N_2^2-1\right) C_{\theta }^2}{3 \epsilon }+O\left(\epsilon ^0\right) \\
(p_6)= & -\frac{8 \left(\pi ^4 N_2 \left(N_1 N_2^2-2 N_2+N_1\right) C_{\theta }^2\right)}{\epsilon }+O\left(\epsilon ^0\right) \\
(p_7)= & \frac{4 \pi ^2 N_2 \left(N_1 \left(N_2^2+3\right)-4 N_2\right) C_{\theta }^2}{\epsilon ^2}+\nonumber\\&\frac{8 \pi ^2 N_2 \left(N_1 \left(N_2^2+3\right)-4 N_2\right) (1+4 \log 2) C_{\theta }^2}{\epsilon }+O\left(\epsilon ^0\right) \\
(p_8)= & \frac{16 \pi ^2 \left(5 \pi ^2-48\right) \left(N_1-N_2\right) N_2 C_{\theta }^2}{3 \epsilon }+O\left(\epsilon ^0\right) \\
(p_9)= & \frac{32 \pi ^2 N_2 \left(N_2-N_1\right) C_{\theta }^2}{\epsilon ^2}-\frac{64 \left(\pi ^2 \left(N_1-N_2\right) N_2 (1+4 \log 2) C_{\theta }^2\right)}{\epsilon }+O\left(\epsilon ^0\right) 
\end{align}

\section{Matrix model correlators}\label{app:matrix}

The perturbative expansion of the Wilson loop is straightforwardly reduced to a combination of correlators in a Gaussian matrix model 
\begin{equation}
Z = \int d \Lambda~ e^{- \alpha \mathrm{Tr}(\Lambda^2)}
\end{equation}
where $\Lambda$ is a $N\times N$ matrix.
The correlators needed for the computation of the four-loop Wilson loop \eqref{eq:Wn} evaluate
\begin{align}
&\left\langle \left(\Tr \Lambda\right)^2\right\rangle =\frac{ N}{2 \alpha },\quad
\left\langle \Tr \Lambda ^2 \right\rangle =\frac{ N^2}{2 \alpha },\quad
\left\langle \left(\Tr \Lambda\right)^2 \Tr \Lambda ^2 \right\rangle =\frac{ N \left( N^2+2\right)}{4 \alpha ^2},
\nonumber\\&
\left\langle \Tr \Lambda ^4 \right\rangle =\frac{2  N^3+ N}{4 \alpha ^2},\quad
\left\langle \left(\Tr \Lambda ^2\right) ^2\right\rangle =\frac{ N^2 \left( N^2+2\right)}{4 \alpha ^2},\quad
\left\langle \left(\Tr \Lambda\right)^4\right\rangle =\frac{3  N^2}{4 \alpha ^2},
\nonumber\\&
\left\langle \Tr \Lambda ^6 \right\rangle =\frac{5  N^2 \left( N^2+2\right)}{8 \alpha ^3},\quad
\left\langle \left(\Tr \Lambda ^2\right) ^3\right\rangle =\frac{ N^2 \left( N^2+2\right) \left( N^2+4\right)}{8 \alpha ^3},\quad
\left\langle \left(\Tr \Lambda\right)^6\right\rangle =\frac{15  N^3}{8 \alpha ^3},
\nonumber\\&
\left\langle \left(\Tr \Lambda\right)^4 \Tr \Lambda ^2 \right\rangle =\frac{3  N^2 \left( N^2+4\right)}{8 \alpha ^3},\quad
\left\langle \left(\Tr \Lambda\right)^2 \left(\Tr \Lambda ^2\right) ^2\right\rangle =\frac{ N \left( N^4+6  N^2+8\right)}{8 \alpha ^3},
\nonumber\\&
\left\langle \Tr \Lambda ^2  \Tr \Lambda ^4 \right\rangle =\frac{ N \left(2  N^4+9  N^2+4\right)}{8 \alpha ^3},\quad
\left\langle \left(\Tr \Lambda\right)^2 \Tr \Lambda ^4 \right\rangle =\frac{ N^2 \left(2  N^2+13\right)}{8 \alpha ^3},\nonumber\\&
\left\langle \Tr \Lambda \Tr \Lambda ^2  \Tr \Lambda ^3 \right\rangle =\frac{3  N^2 \left( N^2+4\right)}{8 \alpha ^3},\quad
\left\langle \Tr \Lambda \Tr \Lambda ^3 \right\rangle =\frac{3  N^2}{4 \alpha ^2},\nonumber\\&
\left\langle \left(\Tr \Lambda\right)^6 \Tr \Lambda ^2 \right\rangle =\frac{15  N^3 \left( N^2+6\right)}{16 \alpha ^4},\quad
\left\langle \left(\Tr \Lambda ^3\right) ^2\right\rangle =\frac{3 \left(4  N^3+ N\right)}{8 \alpha ^3},
\nonumber\\&
\left\langle \left(\Tr \Lambda\right)^2 \left(\Tr \Lambda ^2\right) ^3\right\rangle =
\left\langle \left(\Tr \Lambda ^2\right) ^4\right\rangle =\frac{ N^2 \left( N^6+12  N^4+44  N^2+48\right)}{16 \alpha ^4},\nonumber\\&
\left\langle \left(\Tr \Lambda\right)^3 \Tr \Lambda ^3 \right\rangle =\frac{3  N \left(3  N^2+2\right)}{8 \alpha ^3},\quad
\left\langle \left(\Tr \Lambda\right)^3 \Tr \Lambda ^2  \Tr \Lambda ^3 \right\rangle =\frac{3  N \left(3  N^4+20  N^2+12\right)}{16 \alpha ^4},\nonumber\\&
\left\langle \left(\Tr \Lambda\right)^4 \left(\Tr \Lambda ^2\right) ^2\right\rangle = \left\langle \Tr \Lambda \left(\Tr \Lambda ^2\right) ^2 \Tr \Lambda ^3 \right\rangle =\frac{3  N^2 \left( N^4+10  N^2+24\right)}{16 \alpha ^4},\nonumber\\&
\left\langle \Tr \Lambda ^2  \left(\Tr \Lambda ^3\right) ^2\right\rangle =\frac{3  N \left(4  N^4+25  N^2+6\right)}{16 \alpha ^4},\quad
\left\langle \left(\Tr \Lambda\right)^4 \Tr \Lambda ^4 \right\rangle =\frac{3  N \left(2  N^4+25  N^2+8\right)}{16 \alpha ^4},\nonumber\\&
\left\langle \left(\Tr \Lambda\right)^2 \Tr \Lambda ^2  \Tr \Lambda ^4 \right\rangle =\frac{ N^2 \left(2  N^4+25  N^2+78\right)}{16 \alpha ^4},\nonumber\\&
\left\langle \left(\Tr \Lambda ^2\right) ^2 \Tr \Lambda ^4 \right\rangle =\frac{ N \left(2  N^6+21  N^4+58  N^2+24\right)}{16 \alpha ^4},\quad \left\langle \Tr \Lambda \Tr \Lambda ^5 \right\rangle =\frac{5 \left(2  N^3+ N\right)}{8 \alpha ^3},\nonumber\\&
\left\langle \Tr \Lambda \Tr \Lambda ^3  \Tr \Lambda ^4 \right\rangle =\frac{3  N \left(2  N^4+25  N^2+8\right)}{16 \alpha ^4},\quad
\left\langle \left(\Tr \Lambda ^4\right) ^2\right\rangle =\frac{ N^2 \left(4  N^4+40  N^2+61\right)}{16 \alpha ^4},\nonumber\\&
\left\langle \left(\Tr \Lambda\right)^2 \Tr \Lambda ^6 \right\rangle =\frac{5  N \left( N^4+14  N^2+6\right)}{16 \alpha ^4},\quad
\left\langle \Tr \Lambda ^2  \Tr \Lambda ^6 \right\rangle =\frac{5  N^2 \left( N^4+8  N^2+12\right)}{16 \alpha ^4},\nonumber\\&
\left\langle \Tr \Lambda ^8 \right\rangle =\frac{7  N \left(2  N^4+10  N^2+3\right)}{16 \alpha ^4},\quad
\left\langle \Tr \Lambda \Tr \Lambda ^2  \Tr \Lambda ^5 \right\rangle =\frac{5  N \left(2  N^4+13  N^2+6\right)}{16 \alpha ^4}
\end{align}

\bibliographystyle{JHEP}

\bibliography{biblio}

\end{document}